\documentclass[lettersize,journal]{IEEEtran}
\usepackage{amsmath,amsfonts}
\usepackage{algorithmic}
\usepackage{algorithm}
\usepackage{array}
\usepackage[caption=false,font=normalsize,labelfont=sf,textfont=sf]{subfig}
\usepackage{textcomp}
\usepackage{stfloats}
\usepackage{url}
\usepackage{verbatim}
\usepackage{graphicx}
\usepackage{cite}
\usepackage{xcolor}

\begin{document}

\title{Transiently Driven Reflectionless Resonant Microwave Plasmas via Virtual Critical Coupling}

\author{Muhammad Rizwan Akram,~\IEEEmembership{Member,~IEEE} and Abbas~Semnani,~\IEEEmembership{Senior Member,~IEEE}
\thanks{M. R. Akram is with the Department of Electrical Engineering, School of Electrical Engineering and Computer Science, The Pennsylvania State University, State College, PA 160801, USA (email: rizwanakram373@gmail.com). A. Semnani is with the Department of Electrical Engineering and Computer Science, The University of Toledo, Toledo, Ohio 43606, USA (email: abbas.semnani@utoledo.edu). This work was supported by the Department of Energy (DOE) Grant DE-SC0025743. (Corresponding author: M. R. Akram.)}}

\markboth{IEEE TRANSACTIONS ON Microwave Theory and Techniques,~Vol.~x, No.~xx, 2026}%
{Shell \MakeLowercase{\textit{et al.}}: A Sample Article Using IEEEtran.cls for IEEE Journals}


\maketitle

\begin{abstract}
Microwave plasma sources play a critical role in scientific research and a wide range of industrial, biomedical, and space applications. Resonant microwave structures have recently enabled highly energy-efficient plasma generation by concentrating electromagnetic energy within compact volumes. However, once plasma is ignited, the formation of a conductive region at the resonator's electric-field hotspot significantly perturbs the resonant impedance, resulting in severe impedance mismatch, increased reflection, and reduced power-transfer efficiency. This limitation arises because conventional resonant operation relies on critical coupling, in which the input coupling simultaneously provides impedance matching and perturbs the resonator. This paper overcomes this fundamental limitation by operating the resonator in an over-coupled regime and achieving dynamic impedance matching through temporally modulated excitation. Specifically, an exponentially growing incident waveform is used to emulate the critical coupling condition without physically modifying the resonator, a concept known as virtual critical coupling. The proposed approach enables the resonator to store up to four times as much electromagnetic energy as a conventionally critically coupled resonator. Experimental results demonstrate ultra-efficient resonant microwave plasma generation with multi-fold reductions in ignition energy consumption and enhanced dynamic control over plasma dynamics. 
\end{abstract}


\section{Introduction}
\IEEEPARstart{P}{lasma} sources have become indispensable in a wide range of scientific and engineering applications, including materials processing\cite{r1penkov2015review}, plasma-assisted manufacturing\cite{r2rossnagel1990handbook, r3grill1994cold}, biomedical sterilization\cite{r4laroussi2015low}, propulsion\cite{r5betti2016inertial}, spectroscopy\cite{r6kunze2009introduction}, lighting\cite{r7rakov2013physics}, environmental remediation\cite{r8mizuno2007industrial}, and particle acceleration\cite{r9martinez2013high}. Compared with conventional high-voltage discharge techniques, microwave excitation offers electrode-free operation, reduced contamination, longer device lifetime, and efficient electromagnetic energy delivery to the plasma\cite{r10semnani2016high, r11semnani2016plasma}. Consequently, there has been significant interest in developing compact and energy-efficient microwave plasma sources capable of generating stable plasmas at low input power\cite{r12bardos2010cold}.

Among the various approaches, resonant microwave plasma sources have emerged as an attractive solution because electromagnetic resonance enables strong localization of electric fields within small volumes. The resulting field enhancement substantially reduces the required input power. Numerous resonant structures, including dielectric resonators\cite{r13akram2024anapole, r14akram2024nonradiating}, cavity resonators\cite{r15gulec2015atmospheric}, substrate-integrated waveguides\cite{r16zha2023, r17kabir2025capacitive}, and evanescent-mode resonators\cite{r18sem2022}, have demonstrated efficient plasma generation by exploiting this principle. In particular, atmospheric-air resonant microwave plasma sources have enabled compact plasma jets\cite{r14akram2024nonradiating, r18sem2022} and plasma line sources\cite{r19akram2026energy}, with applications ranging from advanced manufacturing to accelerator technologies\cite{p6wangler2008rf, p7dolgashev2021high, p8tajima1979laser}.

Despite these advantages, resonant microwave plasma sources suffer from a fundamental limitation post-breakdown. Because the plasma is intentionally generated at the electric-field hotspot, the newly formed conductive region significantly perturbs the resonator's electromagnetic properties, altering its resonant frequency, quality factor, and input impedance. As a result, the impedance matching established before breakdown is lost, leading to increased reflection and degraded microwave-to-plasma power-transfer efficiency \cite{r18sem2022, r19akram2026energy}. This post-breakdown impedance mismatch has remained a major obstacle to achieving highly efficient resonant microwave plasma sources, limiting their scalability, plasma density, and dynamic controllability.

The origin of this limitation lies in the operating principle of conventionally excited resonators. Under steady-state excitation, maximum power transfer is typically achieved through critical coupling, in which the input coupling simultaneously excites the resonator and provides conjugate impedance matching. Consequently, any perturbation of the resonator, including the formation of plasma, directly modifies the matching condition and inevitably introduces reflection. Since the input coupling mechanism itself helps establish the impedance match, maintaining efficient power transfer during rapid plasma evolution is fundamentally challenging with conventional excitation techniques.

Recent advances in transient electromagnetic excitation have demonstrated that resonators do not necessarily need to operate exclusively under steady-state conditions \cite{ p16kim2022beyond, p17rasmussen2023lossless}. Concepts based on complex-frequency excitation \cite{p15kim2025complex}, time-reversed waveforms\cite{q0kalluri2018electromagnetics, q1lerosey2004time, q2mazieres2019plasma, q3mazieres2021space}, and virtual critical coupling\cite{p20delage2022experimental,p21krasnok2019anomalies,p22hinney2024efficient} have shown that appropriately designed transient signals can dramatically modify the exchange of electromagnetic energy between a source and a resonator. In particular, virtual critical coupling\cite{q4delage2023plasma, q5marini2022perfect} employs an exponentially growing incident waveform to emulate the impedance-matching condition of a critically coupled resonator while physically operating the device in an over-coupled regime. This approach enables nearly reflectionless transient excitation and allows substantially greater electromagnetic energy to accumulate within the resonator than is possible under conventional critical coupling.

In this work, we extend the concept of virtual critical coupling to our resonant microwave plasma sources and demonstrate a general approach to reflectionless and energy-efficient plasma generation. By transiently exciting an over-coupled resonator with customized exponentially growing microwave bursts, efficient electromagnetic energy storage is maintained throughout plasma formation while minimizing reflected power. The proposed technique is experimentally demonstrated with an anapole dielectric resonator that operates as a resonant microwave plasma source. The presented approach significantly reduces the energy required for plasma ignition, improves microwave-to-plasma energy transfer efficiency, and enables enhanced dynamic control over plasma formation.

The main contributions of this paper are threefold. First, virtual critical coupling enables up to a fourfold increase in the electromagnetic energy stored within the resonator compared with conventional critical coupling, substantially lowering the plasma breakdown threshold. Second, reflectionless transient excitation enables stable plasma generation with lower energy consumption than conventional resonant excitation. Third, sustained reflectionless energy delivery improves plasma density and enhances dynamic control of the generated plasma, establishing a general framework for next-generation, high-efficiency resonant microwave plasma sources applicable across a wide range of microwave plasma technologies.

\section{Theory and Approach}
Starting with a brief overview of the theory of complex frequency excitations \cite{p15kim2025complex,p16kim2022beyond,p17rasmussen2023lossless}, the dynamic equation for a resonant mode of a single-port system can be written as follows \cite{p18fan2003temporal}
\begin{equation}\label{Eq.1}
    \frac{da}{dt} = (j\omega_0 - \gamma_{int} - \gamma_{ext})a + \kappa S^{inc},
\end{equation}
where $\gamma_{int}$ is the loss internally in the cavity, $\gamma_{ext}$ is the leakage of energy to the ports. $\omega_0$ is the resonant frequency, $\kappa$ is the coupling with the incident port, and $a$ is the mode amplitude. The mode amplitude evolves over time with $e^{-j\omega_0 t}e^{-\gamma t}$, where the first factor is sinusoidal and the second factor is exponential decay. The reflection from the resonant mode can be described as
\begin{equation}\label{Eq.2}
    S^{ref} = C S^{inc} + a \kappa,
\end{equation}
where $C$ is the direct coupling between the ports and $S^{ref}$ is the reflection from the resonant mode to the port. For a single-port system, $C$ can be replaced by -1, giving $S^{ref} = -S^{inc} + a\kappa$. Taking the excitation signal as $S^{inc} ~ e^{j\omega t}$, (\ref{Eq.1}) can be rewritten in the frequency domain as $a = kS^{inc}/[j(\omega - \omega_0) +\gamma_{int} + \gamma_{ext})]$. From these two expressions and the expression of $k = \sqrt{2\gamma_{ext}}$, we can derive the reflection amplitude in terms of the excitation frequency $\omega= \omega+j\omega''$ as $(r =S^{ref}/S^{inc})$ in the complex plane as\cite{p19ra2020virtual, p20delage2022experimental}
\begin{equation}\label{equ. 3}
    r(\omega', \omega'') =  \frac{(\gamma_{ext} -\gamma_{int}+\omega'')-j(\omega'-\omega_0)}{(\gamma_{ext} +\gamma_{int}-\omega'')+j(\omega'-\omega_0)}.
\end{equation}

\begin{figure}[!b]
\centering
\includegraphics[width=0.98\linewidth]{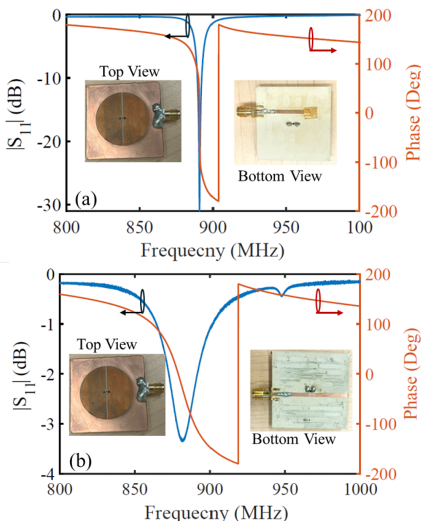}
\caption{\label{fig:fig1} (a) Magnitude and phase response of the reflection coefficient ($S_{11}$) for the critically coupled anapole resonator for a single plasma jet, with top/bottom views showing the resonator/feeding board in the insets. (b) Magnitude and phase response of the reflection coefficient for the over-coupled anapole resonator with a single plasma jet; top- and bottom-view images of the resonator/feeding board are shown in the insets.}
\end{figure}

For a port system, the reflection amplitude as obtained in (\ref{equ. 3}) corresponds to the scattering matrix \cite{p21krasnok2019anomalies} of the resonator through which the resonant behavior can be fully understood. The pole of the S-matrix indicates the decay of the stored energy through the reflection channel and the internal losses (radiative/non-radiative) in the absence of excitation. It is described by the complex frequency $\omega = \omega'+j\omega'' = \omega_0+j(\gamma_{int}+\gamma_{ext})$. The zero of the S matrix as obtained from (\ref{equ. 3}) demonstrates the condition for complete suppression of the reflection if it is excited with the complex frequency of $\omega = \omega'+j\omega'' = \omega_0 + j(\gamma^{ext}-\gamma^{int})$. The special case($\gamma^{ext}=\gamma^{int}$) refers to critically coupled systems in which sinusoidal excitation is employed to transfer the power to the system with minimized reflection after a transient period. Critical coupling is achieved by introducing coupling loss, thereby reducing the resonator's overall energy-storage capacity.

To enhance and utilize the resonator's energy-storage capacity, it is important to account for the temporal evolution of energy. The expression for the energy can be obtained from the resonance amplitude in terms of the quality factor and decay rates as
\begin{equation}\label{eq.4}
    |a|^2 = \frac{4Q_{ext}}{\omega_0}[\frac{\gamma_{ext}^2}{(\gamma+(\omega-\omega_0))^2}]|S^{inc}|^2,
\end{equation}
 where $\gamma = \gamma_{int}+\gamma_{ext}$ and $Q_{ext} = \omega_0/2\gamma_{ext}$. Then, (\ref{eq.4}) can be written in terms of quality factors as\cite{p19ra2020virtual}
 
\begin{equation}\label{eq.5}
    |a|^2 = \frac{4}{\omega_0}[\frac{Q_{ext}}{(1+Q_{ext}/Q_{int})^2}]|S^{inc}|^2,
\end{equation}
where $Q_{int} = \omega_0/2\gamma_{int}$ is the intrinsic quality factor of the resonator, which depends on the radiative and internal losses of the cavity. For a critically coupled system, the excitation is pure sinusoidal, i.e., $e^{-\omega t}$ and $Q_{ext} = Q_{int}$, the  maximum stored energy can be evaluated as
\begin{equation}\label{eq.6}
    |a|^2 = \frac{Q_{ext}}{\omega_0} S_0^2 = U^cS_0^2,
\end{equation}
where $U^c$ is the maximum stored energy, scaled by the incident signal energy in the case of the monochromatic wave at the resonance of the critically coupled system. On the other hand, when the resonant mode/cavity is excited in such a way that $Q_{ext}<<Q_{int}$ implies $\gamma_{int} << \gamma_{ext}$, the resonator is said to be over-coupled. Then, (\ref{eq.6}) can be evaluated as follows
\begin{equation}\label{eq.7}
    |a|^2 = \frac{4Q_{ext}}{\omega_0}|S^{inc}|^2 = 4U^c|S^{inc}|^2.
\end{equation}
\begin{figure*}[!b]
\centering
\includegraphics[width=0.9\linewidth]{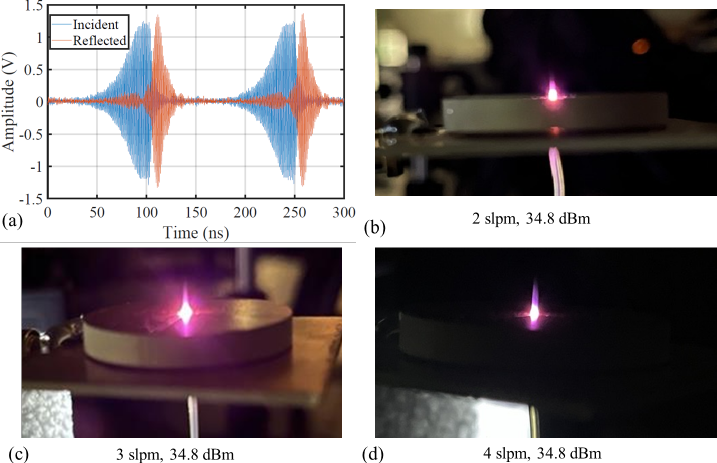}
\caption{\label{fig:fig2}(a) A snippet of the designed excitation signal at $\omega_z = 882.98+j10$ MHz (blue), repeating indefinitely and the reflection at the port (orange) in response to the excitation signal, and plasma jet at 34.8 dBm of input power for helium flow rate of (b) 2, (c) 3, and (d) 4 slpm.}
\end{figure*}
The relation in (\ref{eq.7}) clearly demonstrates in over-coupled scenarios, i.e., $\gamma^{int} <<\gamma^{ext}$, that it is possible to scale the energy up to 4 times for monochromatic signals and for complex signals $4|S^{inc}|^2$. By designing the over-coupled resonator and applying the excitation signal at zeros of the S matrix of (\ref{equ. 3}), one can practically realize almost reflectionless cavity excitation with infinite storage capacity. Excitation through zero of the system is known as virtual critical coupling \cite{p19ra2020virtual}. However, excitation through zero typically requires an input signal that is exponentially growing, such as $S^{inc} \propto e^{-\omega t}e^{\gamma^{ext}t}$. This would potentially limit the maximum power delivered to the cavity. In the remainder of the paper, we will explore how to leverage this enhanced energy-storage capability to enable atmospheric plasma line jets for applications in plasma wakefield accelerators and, more broadly, in propulsion, fusion, surface treatment, biomedical, and agricultural domains.

\section{Results and Discussion}
To demonstrate the concept, we have utilized the anapole resonator as proposed in our earlier papers. The anapole resonator is operated in two distinct coupling regimes, as explained in the earlier section: (1) a conventional critical coupling regime, and (2) the proposed over-coupled regime. In the critical coupling regime, the energy is coupled to the anapole resonator via a microstrip line, such that after reaching steady state, the device reflects very little at the input port for a monochromatic wave. The reflection coefficient of the device is shown in Fig. \ref{fig:fig1}(a), which indicates reflection below -30 dB at a resonance frequency of 890 MHz. This is achieved by positioning the microstrip line relative to the resonator's coupling slot. For the over-coupling regime, the coupling of the microstrip line is adjusted by placing it with respect to the slot and the length of the microstrip in such a way as to ensure $\gamma^{int} <<\gamma^{ext}$. The over-coupling regime can be verified through the reflection coefficient as presented in Fig. \ref{fig:fig1}(b), where a reflection coefficient close to 0 dB is desired with a phase transition from -180$^o$ to +180$^o$.

To minimize reflection for the over-coupled resonator, it should be excited at zero frequency, $\omega_z$. To obtain $\omega_z$, the time domain response of the reflection of the device is recorded with an oscilloscope under monochromatic wave excitation at the center frequency of 880 MHz. We observe the cancellation of incident and reflected waves at $\tau_0=$11 ns. The reflection amplitude in steady state $r_\infty$ = 0.8289, both of which are crucial to determine the envelope of the sinusoidal signal given by the imaginary part of $\omega_z$ \cite{p20delage2022experimental,p22hinney2024efficient}. The real part of $\omega_z$ can be obtained by the Fourier transform of the reflecting signal at the port as soon as the input is cut off. Using $\omega_z$, the anapole resonator is then excited to realize a perfect matching condition for the entire duration of the exponential growth input given by $e^{-i\omega_zt}$. Practically, it is only feasible to maintain the exponential growth for short durations, which usually depend on the maximum power available and the envelope frequency, i.e., the imaginary part of $\omega_z$.

\begin{figure}[!]
\centering
\includegraphics[width=0.95\linewidth]{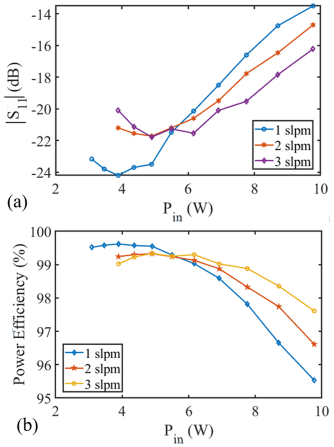}
\caption{\label{fig:Fig3} Single plasma jet using a critically coupled resonator: (a) magnitude of the reflection coefficient, and (b) power coupling efficiency.}
\end{figure}

The power is delivered to the resonator via the designed excitation signal, which has a pulse duration of approximately 145 ns and repeats indefinitely. Helium is pushed through the anapole resonator with a controlled flow rate through a central cylindrical hole inside the resonator. The power is gradually increased until the gas breaks down into a plasma and is driven out of the resonator by the high gas velocity in the narrow channel. The presence of the conductive region within the enhanced electric-field region will slightly perturb the resonance. To avoid reflection due to a mismatch caused by the presence of plasma, the excitation signal is tailored in the presence of plasma and is given by $\omega_z = 882.98+j10$ MHz. The incident and reflected signals at the resonator port are plotted in Fig. \ref{fig:fig2}(a). It is observed that there is very low reflection as long as the incident signal is applied, followed by the release of energy as the reflected power at the port when the input pulse terminates.

\begin{figure}[!]
\centering
\includegraphics[width=0.95\linewidth]{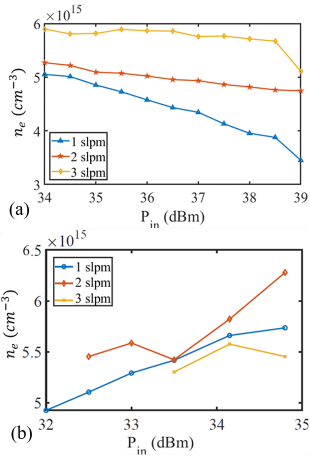}
\caption{\label{fig:Fig3a} Comparison of electron density characteristics versus $P_{in}$ for (a) a critically coupled resonator under monochromatic excitation, and (b) an over-coupled resonator under complex excitation.}
\end{figure}

The plasma jet realized in response to the excitation signal is shown in Figs. \ref{fig:fig2}(b-d) at various flow rates of helium. The gas breakdown occurs at 34.2 dBm (2.5 W) compared to 36 dBm (4 W) for the case of critical coupling when measured using the same anapole resonator but different bottom feeding boards as shown in Fig. \ref{fig:fig1}, bottom view in the insets. The input power is measured using Keysight power sensors. About 1.8 dB less power is required to generate plasma in the virtual critical coupling case than in the critical coupling scenario. To further assess the impact of the two distinct operating regimes of the resonator on the resulting plasma, the plasma composition (i.e., electron density) was measured using a spectrometer. The electron density evaluated is plotted in Figs. \ref{fig:Fig3a}(a,b).

\begin{figure}[!]
\centering
\includegraphics[width=0.95\linewidth]{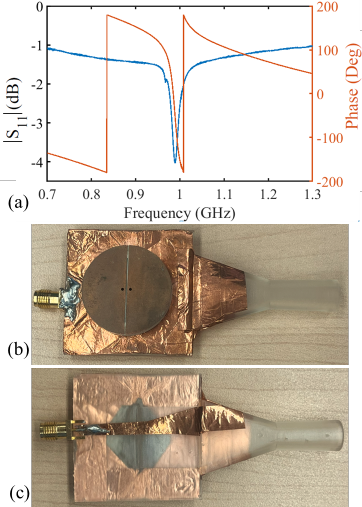}
\caption{\label{fig:Fig4} (a) Magnitude and phase of the reflection coefficient for the over-coupled anapole resonator for a 2-cm line plasma jet. (b) Top and (c) bottom views of the assembled device.}
\end{figure}

The general trend in the electron density of the plasma jet is upward for the virtual coupling regime and downward for the critical coupling regime. This is mainly because, in the virtual critical coupling regime, the maximum power delivered to the resonator is limited by the available power amplifier's upper bound, which is typically 10-12 dB lower than in the critical coupling regime due to the additional circuitry required for modulation. To further assess why the electron density is decreasing with increasing input power, the power coupling efficiency and the reflection response are plotted in Figs. \ref{fig:Fig3}(a,b). As power is increased, we observe that power coupling efficiency drops by 4-5$\%$ and $S_{11}$ increases from -24 to -14 dB. The second observation, consistent with what we have discussed so far, is that as the flow rate increases, the electron density in the critical-coupling case becomes flatter. This is primarily because the plasma is displaced farther from the intense-field region by the higher flow rate.

\begin{figure}[!]
\centering
\includegraphics[width=0.95\linewidth]{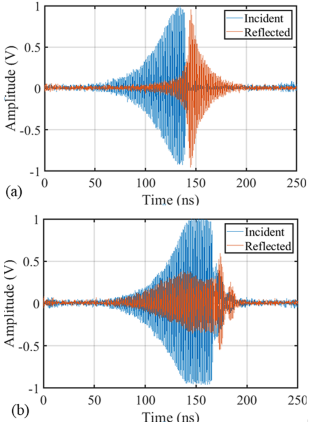}
\caption{\label{fig:Fig5} A snippet of the designed excitation signal at $\omega_z = 900+j7$ MHz (blue), repeating indefinitely, and the reflection at the port (orange) in response to the excitation signal: (a) pre-breakdown and (b) post-breakdown scenarios.}
\end{figure}

To further evaluate the effectiveness of the proposed approach, the anapole device for the plasma jet line, as proposed in \cite{r19akram2026energy}, was modified to operate in an over-coupled scenario, as can be observed in Fig. \ref{fig:Fig4}(a). The top and bottom views of the device are shown in Fig. \ref{fig:Fig4}(b) and (c). The modification is intended solely for the feeding board, specifically to adjust the microstrip dimensions. The next step is to design an appropriate excitation for the over-coupled anapole resonator in a line jet. Using monochromatic excitation in pulse mode, the time instant at which the input completely cancels the reflected signal, i.e., $\tau_0=18$ ns, is evaluated, and the steady state amplitude $r_\infty$=0.65 is measured in line with what is obtained from $S_{11}$. Using these measurements, the excitation frequency is obtained as $\omega_0 = 900+j7$ MHz. The designed pulse and the reflected signal from the anapole resonator are shown in Fig. \ref{fig:Fig5} for scenarios without and with plasma. The 185 ns pulse is repeated to maintain continuous plasma formation. Without plasma, the device maintains near-reflectionless coupling to the excitation signal, and energy is released as the input approaches zero. After plasma formation, the device is no longer operating in an over-coupled regime. Still, the device performs better than in the critical coupling regime, where impedance matching is disrupted after plasma formation, resulting in reflected power exceeding 90$\%$.

\begin{figure*}[!]
\centering
\includegraphics[width=0.9\linewidth]{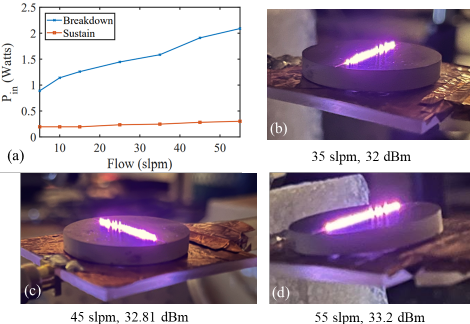}
\caption{\label{fig:Fig6} (a) Breakdown and sustaining powers for the 2-cm plasma jet line under virtual critical coupling. Helium flow rate of (b) 35 slpm at 32 dBm input power, (c) 45 slpm at 32.81 dBm input power, and (d) 55 slpm at 33.2 dBm input power.}
\end{figure*}

The realized plasma jet lines are shown in Figs. \ref{fig:Fig6}(b-d) at three different helium flow rates. The plasma jets formed are very uniform across the device. The small discontinuities are mainly caused by the construction of the channel through the device, as discussed in \cite{r19akram2026energy}. It is interesting to note that the power required to form a uniform plasma line jet ranges from 32 dBm (1.58 W) to 33.2 dBm (2.08 W) for helium flow rates of 1-55 slpm, which is a fraction of the power required when the device is operating in critical coupling mode. Typically, at least 25 W of power is required when the device operates in the critical-coupling regime to maintain the plasma jet in the line at higher flow rates, e.g., 55 slpm \cite{r19akram2026energy}. The proposed approach of virtually critical coupling enabled the realization of plasma jet lines using anapole resonators, not only because of the very low input power but also because it made it possible to directly ignite uniform plasma line jets at higher flow rates. The power required for the line jet is lower than that for a single jet. It is mainly caused by two reasons: (1) a lower electric field in the center of the device, and (2) a large outlet of 0.6 mm diameter for the single jet and 0.2 mm for the line jet; the lower gap requires low voltage for ignition, as obtained from E = V/d.

\section{Conclusion}
This work successfully demonstrated a novel, fully planar, compact, and frequency-tunable atmospheric pressure plasma jet device. This plasma jet technology leverages the capabilities of a dielectric anapole structure, a non-radiating resonator, to enhance the near electric field while effectively suppressing far-field radiation. This demonstration represents a pivotal step toward a new pathway to highly efficient plasma sources with minimal reflection and radiation. The key advantages of the proposed technology, including its high electron density, compact form factor, seamless integration capability, and cost-effectiveness, hold the potential to open up new horizons for discovery and application. Given the profound impact of plasma technology across many fields, these advanced attributes are positioned to make substantial contributions to progress in various domains. Moreover, the inherent ease of frequency tunability of the proposed technology holds promise for exploring enriched chemistry within the RF/Microwave spectrum, further enhancing its utility and versatility.



\bibliographystyle{IEEEtran}
\bibliography{ref}

\begin{thebibliography}{10}
\providecommand{\url}[1]{#1}
\csname url@samestyle\endcsname
\providecommand{\newblock}{\relax}
\providecommand{\bibinfo}[2]{#2}
\providecommand{\BIBentrySTDinterwordspacing}{\spaceskip=0pt\relax}
\providecommand{\BIBentryALTinterwordstretchfactor}{4}
\providecommand{\BIBentryALTinterwordspacing}{\spaceskip=\fontdimen2\font plus
\BIBentryALTinterwordstretchfactor\fontdimen3\font minus \fontdimen4\font\relax}
\providecommand{\BIBforeignlanguage}[2]{{%
\expandafter\ifx\csname l@#1\endcsname\relax
\typeout{** WARNING: IEEEtran.bst: No hyphenation pattern has been}%
\typeout{** loaded for the language `#1'. Using the pattern for}%
\typeout{** the default language instead.}%
\else
\language=\csname l@#1\endcsname
\fi
#2}}
\providecommand{\BIBdecl}{\relax}
\BIBdecl

\bibitem{r1penkov2015review}
O.~V. Penkov, M.~Khadem, W.-S. Lim, and D.-E. Kim, ``A review of recent applications of atmospheric pressure plasma jets for materials processing,'' \emph{Journal of Coatings Technology and Research}, vol.~12, no.~2, pp. 225--235, 2015.

\bibitem{r2rossnagel1990handbook}
S.~M. Rossnagel, J.~J. Cuomo, and W.~D. Westwood, ``Handbook of plasma processing technology: fundamentals, etching, deposition, and surface interactions,'' \emph{(No Title)}, 1990.

\bibitem{r3grill1994cold}
A.~Grill, \emph{Cold plasma in materials fabrication}.\hskip 1em plus 0.5em minus 0.4em\relax IEEE Press, New York, 1994, vol. 151.

\bibitem{r4laroussi2015low}
M.~Laroussi, ``Low-temperature plasma jet for biomedical applications: a review,'' \emph{IEEE transactions on plasma science}, vol.~43, no.~3, pp. 703--712, 2015.

\bibitem{r5betti2016inertial}
R.~Betti and O.~Hurricane, ``Inertial-confinement fusion with lasers,'' \emph{Nature Physics}, vol.~12, no.~5, pp. 435--448, 2016.

\bibitem{r6kunze2009introduction}
H.-J. Kunze, \emph{Introduction to plasma spectroscopy}.\hskip 1em plus 0.5em minus 0.4em\relax Springer Science \& Business Media, 2009, vol.~56.

\bibitem{r7rakov2013physics}
V.~Rakov, ``The physics of lightning,'' \emph{Surveys in Geophysics}, vol.~34, no.~6, pp. 701--729, 2013.

\bibitem{r8mizuno2007industrial}
A.~Mizuno, ``Industrial applications of atmospheric non-thermal plasma in environmental remediation,'' \emph{Plasma Physics and Controlled Fusion}, vol.~49, no.~5A, pp. A1--A15, 2007.

\bibitem{r9martinez2013high}
A.~Martinez de~la Ossa, J.~Grebenyuk, T.~Mehrling, L.~Schaper, and J.~Osterhoff, ``High-quality electron beams from beam-driven plasma accelerators<? format?> by wakefield-induced ionization injection,'' \emph{Physical review letters}, vol. 111, no.~24, p. 245003, 2013.

\bibitem{r10semnani2016high}
A.~Semnani, S.~O. Macheret, and D.~Peroulis, ``A high-power widely tunable limiter utilizing an evanescent-mode cavity resonator loaded with a gas discharge tube,'' \emph{IEEE Transactions on Plasma Science}, vol.~44, no.~12, pp. 3271--3280, 2016.

\bibitem{r11semnani2016plasma}
A.~Semnani, D.~Peroulis, and S.~O. Macheret, ``Plasma-enabled tuning of a resonant rf circuit,'' \emph{IEEE transactions on Plasma Science}, vol.~44, no.~8, pp. 1396--1404, 2016.

\bibitem{r12bardos2010cold}
L.~B{\'a}rdos and H.~Bar{\'a}nkov{\'a}, ``Cold atmospheric plasma: Sources, processes, and applications,'' \emph{Thin solid films}, vol. 518, no.~23, pp. 6705--6713, 2010.

\bibitem{r13akram2024anapole}
M.~R. Akram and A.~Semnani, ``Anapole source based on electric dipole interactions over a low-index dielectric,'' \emph{Physical Review Applied}, vol.~21, no.~5, p. 054051, 2024.

\bibitem{r14akram2024nonradiating}
------, ``Nonradiating resonances: Anapoles enabling highly efficient plasma jets within dielectric structures,'' \emph{IEEE Transactions on Microwave Theory and Techniques}, 2024.

\bibitem{r15gulec2015atmospheric}
A.~Gulec, F.~Bozduman, and A.~M. Hala, ``Atmospheric pressure 2.45-ghz microwave helium plasma,'' \emph{IEEE Transactions on Plasma Science}, vol.~43, no.~3, pp. 786--790, 2015.

\bibitem{r16zha2023}
C.~Zhao, X.~Li, D.~K. Agrawal, Z.~Yan, S.~Qi, Y.~Liu, T.~Ma, Q.~Chen, Y.~Zhang, C.~Wang \emph{et~al.}, ``Microwave atmospheric pressure plasma jet generated from substrate integrated waveguide resonator,'' \emph{Plasma Processes Polymers}, p. e2200230, 2023.

\bibitem{r17kabir2025capacitive}
K.~S. Kabir, K.~Singhal, and A.~Semnani, ``Capacitive-tuned siw evanescent-mode cavity for resonant microwave plasma jets,'' \emph{IEEE Transactions on Microwave Theory and Techniques}, 2025.

\bibitem{r18sem2022}
A.~Semnani and K.~S. Kabir, ``A highly efficient microwave plasma jet based on evanescent-mode cavity resonator technology,'' \emph{IEEE Trans. Plasma Sci.}, vol.~50, no.~10, pp. 3516--3524, 2022.

\bibitem{r19akram2026energy}
M.~R. Akram and A.~Semnani, ``An energy-efficient atmospheric plasma jet line enabled by a dielectric microwave anapole source,'' \emph{IEEE Transactions on Plasma Science}, 2026.

\bibitem{p6wangler2008rf}
T.~P. Wangler, \emph{RF Linear accelerators}.\hskip 1em plus 0.5em minus 0.4em\relax John Wiley \& Sons, 2008.

\bibitem{p7dolgashev2021high}
V.~Dolgashev, L.~Faillace, B.~Spataro, S.~Tantawi, and R.~Bonifazi, ``High-gradient rf tests of welded x-band accelerating cavities,'' \emph{Physical Review Accelerators and Beams}, vol.~24, no.~8, p. 081002, 2021.

\bibitem{p8tajima1979laser}
T.~Tajima and J.~M. Dawson, ``Laser electron accelerator,'' \emph{Physical review letters}, vol.~43, no.~4, p. 267, 1979.

\bibitem{p16kim2022beyond}
S.~Kim, S.~Lepeshov, A.~Krasnok, and A.~Al{\`u}, ``Beyond bounds on light scattering with complex frequency excitations,'' \emph{Physical Review Letters}, vol. 129, no.~20, p. 203601, 2022.

\bibitem{p17rasmussen2023lossless}
C.~Rasmussen, M.~I. Rosa, J.~Lewton, and M.~Ruzzene, ``A lossless sink based on complex frequency excitations,'' \emph{Advanced Science}, vol.~10, no.~28, p. 2301811, 2023.

\bibitem{p15kim2025complex}
S.~Kim, A.~Krasnok, and A.~Al{\`u}, ``Complex-frequency excitations in photonics and wave physics,'' \emph{Science}, vol. 387, no. 6741, p. eado4128, 2025.

\bibitem{q0kalluri2018electromagnetics}
D.~K. Kalluri, \emph{Electromagnetics of time varying complex media: frequency and polarization transformer}.\hskip 1em plus 0.5em minus 0.4em\relax CRC Press, 2018.

\bibitem{q1lerosey2004time}
G.~Lerosey, J.~de~Rosny, A.~Tourin, A.~Derode, G.~Montaldo, and M.~Fink, ``Time reversal of electromagnetic waves,'' \emph{Physical review letters}, vol.~92, no.~19, p. 193904, 2004.

\bibitem{q2mazieres2019plasma}
V.~Mazi{\`e}res, R.~Pascaud, L.~Liard, S.~Dap, R.~Clergereaux, and O.~Pascal, ``Plasma generation using time reversal of microwaves,'' \emph{Applied Physics Letters}, vol. 115, no.~15, 2019.

\bibitem{q3mazieres2021space}
V.~Mazi{\`e}res, O.~Pascal, R.~Pascaud, L.~Liard, S.~Dap, R.~Clergereaux, and J.-P. Boeuf, ``Space-time plasma-steering source: Control of microwave plasmas in overmoded cavities,'' \emph{Physical Review Applied}, vol.~16, no.~5, p. 054038, 2021.

\bibitem{p20delage2022experimental}
T.~Delage, O.~Pascal, J.~Sokoloff, and V.~Mazi{\`e}res, ``Experimental demonstration of virtual critical coupling to a single-mode microwave cavity,'' \emph{Journal of Applied Physics}, vol. 132, no.~15, 2022.

\bibitem{p21krasnok2019anomalies}
A.~Krasnok, D.~Baranov, H.~Li, M.-A. Miri, F.~Monticone, and A.~Al{\'u}, ``Anomalies in light scattering,'' \emph{Advances in Optics and Photonics}, vol.~11, no.~4, pp. 892--951, 2019.

\bibitem{p22hinney2024efficient}
J.~Hinney, S.~Kim, G.~J. Flatt, I.~Datta, A.~Al{\`u}, and M.~Lipson, ``Efficient excitation and control of integrated photonic circuits with virtual critical coupling,'' \emph{Nature Communications}, vol.~15, no.~1, p. 2741, 2024.

\bibitem{q4delage2023plasma}
T.~Delage, J.~Sokoloff, O.~Pascal, V.~Mazi{\`e}res, A.~Krasnok, and T.~Callegari, ``Plasma ignition via high-power virtual perfect absorption,'' \emph{ACS photonics}, vol.~10, no.~10, pp. 3781--3788, 2023.

\bibitem{q5marini2022perfect}
A.~V. Marini, D.~Ramaccia, A.~Toscano, and F.~Bilotti, ``Perfect matching of reactive loads through complex frequencies: From circuital analysis to experiments,'' \emph{IEEE Transactions on Antennas and Propagation}, vol.~70, no.~10, pp. 9641--9651, 2022.

\bibitem{p18fan2003temporal}
S.~Fan, W.~Suh, and J.~D. Joannopoulos, ``Temporal coupled-mode theory for the fano resonance in optical resonators,'' \emph{Journal of the Optical Society of America A}, vol.~20, no.~3, pp. 569--572, 2003.

\bibitem{p19ra2020virtual}
Y.~Ra’di, A.~Krasnok, and A.~Al{\'u}, ``Virtual critical coupling,'' \emph{ACS photonics}, vol.~7, no.~6, pp. 1468--1475, 2020.

\end{thebibliography}

\newpage

 




\vfill

\end{document}